%% file: main.tex
\documentclass[conference]{IEEEtran}
\newif\iflinks
\linkstrue

\usepackage{endnotes}

\usepackage{graphicx}  
\usepackage{verbatim}  
\usepackage{latexsym}  
\usepackage{amsmath,amssymb}   
\usepackage{upgreek}
\usepackage{subfigure} 
\usepackage{cite}      
\usepackage[table,dvipsnames]{xcolor} 
\usepackage[nomessages]{fp} 
\usepackage{booktabs}

\iflinks
\usepackage[hypertexnames=true]{hyperref} 
\fi

\newif\ifdraft

\newif\ifbrief

\newcommand{\tsqrd}{\textsuperscript{2}}
\newcommand{\tss}[1]{\textsubscript{#1}}
\newcommand{\tps}[1]{\textsuperscript{#1}}

\colorlet{CmntClr}{black!60!green}

\colorlet{EditClr}{black!10!orange}

\colorlet{OkGreen}{blue!30!black!60!green}

\newlength{\lcbox}
\iflinks
\hypersetup{
	colorlinks,
	linkcolor={red!75!black},
	citecolor={blue!60!black},
	urlcolor={blue!80!black}
}
\fi

\newcommand{\NN}{{\sf I\kern-0.14emN}}   
\newcommand{\ZZ}{{\sf Z\kern-0.45emZ}}   
\newcommand{\QQQ}{{\sf C\kern-0.48emQ}}   
\newcommand{\RR}{{\sf I\kern-0.14emR}}   

\newcommand{\rDelta}{$\Updelta$}
\newcommand{\rSigma}{$\Upsigma$}

\newcommand{\rDS}{$\Updelta\Upsigma$}

\newcommand{\rmu}{$\upmu$}

\newcommand{\rsigma}{$\upsigma$}
\newcommand{\rtau}{$\uptau$}

\newcommand{\syncc}{~\stackrel{\textstyle \rhd\kern-0.57em\lhd}{\scriptstyle L}~}

\begin{document}
\bstctlcite{IEEEexample:BSTcontrol}
\title{A 68\rmu W 31\,kS/s Fully-Capacitive Noise-Shaping SAR ADC with 102\,dB SNDR}
\author{Lieuwe B. Leene\textsuperscript{$\dagger$ $\star$}, Shiva Letchumanan\textsuperscript{$\dagger$ $\star$}, Timothy G. Constandinou\textsuperscript{$\dagger$ $\star$}%
\\$^\star$Centre for Bio-Inspired Technology, Institute of Biomedical Engineering, Imperial College London, SW7 2AZ, UK%
\\$^\dagger$Department of Electrical and Electronic Engineering, Imperial College London, SW7 2AZ, UK%
\\Email: \{l.leene, t.constandinou\}@imperial.ac.uk\vspace{-5pt}}%

\maketitle

\renewcommand\thefootnote{}\thefootnote{}\footnote{This preprint is the “accepted” version by IEEE ISCAS$\copyright$2019 IEEE. Personal use of this material is permitted. Permission from IEEE must be obtainedfor  all  other  uses,  in  any  current  or  future  media,  including  reprinting/republishing  this  material  for  advertising  or  promotional  purposes,  creating  newcollective works, for resale or redistribution to servers or lists, or reuse of any copyrighted component of this work in other works.}

\begin{abstract} \label{Sec:Abstract}
This paper presents a 17\,bit analogue-to-digital converter that incorporates mismatch and quantisation noise-shaping techniques into an energy-saving 10\,bit successive approximation quantiser to increase the dynamic range by another 42\,dB. We propose a novel fully-capacitive topology which allows for high-speed asynchronous conversion together with a background calibration scheme to reduce the oversampling requirement by 10$\times$ compared to prior-art. A 0.18\rmu m CMOS technology is used to demonstrate preliminary simulation results together with analytic measures that optimise parameter and topology selection. The proposed system is able to achieve a FoM\tss{S} of 183\,dB for a maximum signal bandwidth of 15.6\,kHz while dissipating 68\,\rmu W from a 1.8\,V supply. A peak SNDR of 102\,dB is demonstrated for this rate with a 0.201\,mm\tsqrd \:area requirement.
\end{abstract}

\section{Introduction} \label{Sec:Introduction}
Analogue-to-digital converter (ADC) efficiency remains to be the highlight for many current developments in both industry and academia. It used to be the case that oversampling converters (\rDelta\rSigma~ADCs) and successive-approximation register converters (SAR ADCs) found separate application domains where this factor peaks. State-of-the-art ADCs however have mixed these two digitisation techniques to improve performance beyond a 170\,dB Schreier Figure-of-Merit (FoM\tss{S})\cite{nssar3,nssar1,nssar4,nssar5,nssar6}. This trend is in-part driven by the growing bio-metric and bio-medical electronics market that necessitates low-power high dynamic-range signal acquisition as many phenomena of interest exhibit signal dynamics with several orders of magnitude in variation. For example a peripheral neuro-modulation device with digitally assisted artifact rejection\cite{dneuromod} requires over $>$100\,dB of dynamic range to detect micro-volt level sensory neuron activity in the presence of large mili-volt level interference from stimulation or motor-unit activity which is the application of interest that motivated this work.

The emerging ADC topologies for bio-sensors use multi-stage noise shaping or pipe-lined operation where multiple quantisers are integrated together and the quantisation error of the first quantiser is either resolved by another quantiser after amplification or may be used directly with an alternate feedback mechanism to similarly resolve additional bits. The noise-shaping SAR (NS-SAR) \cite{nssar0,nssar1} however adopts a different approach by sampling and converting the input multiple times while simultaneously employing multiple feedback mechanisms that up-modulate any conversion errors out of the signal bandwidth. In this way the signal can be resolved with much finer precision once the output is decimated and the out-of-band frequency components are filtered out.

\begin{figure}
	\centering
	\includegraphics[width=7cm]{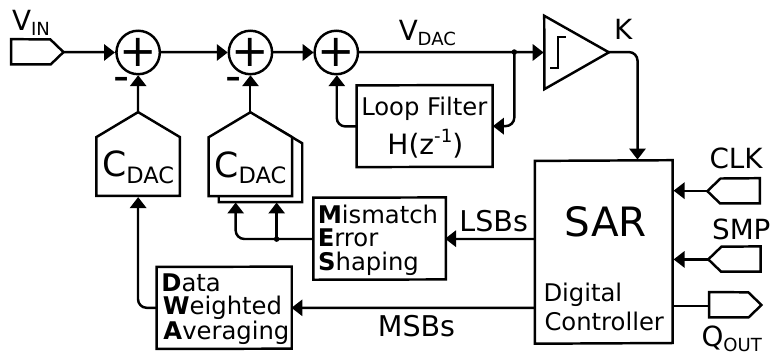}
	\caption{Block diagram of the proposed high-resolution data converter showing the SAR digital controller applying feedback through 3 separate capacitor arrays and is augmented by the switched-capacitor loop filter H(z\tps{-1}).}
	\label{Fig:concept}
	\vspace{-5mm}
\end{figure}%

Here we present a novel fully-capacitive NS-SAR topology using active higher-order noise shaping that achieves state-of-the-art efficiency for high resolution signal acquisition. The proposed configuration is shown in Fig.~\ref{Fig:concept}. This figure summarises which signals are processed by each block in a closed-loop fashion to resolve the sampled analogue input signal V\tss{IN}. The main data-conversion mechanism is based on the conventional SAR controller that uses the comparator decisions K to successively set the MSB and LSB bits\cite{saradc1}. However to augment this operation two separate noise-shaping mechanisms are added; one for quantisation noise, H(z\tps{-1}), and another for mismatch noise by means of data-weighted averaging (DWA) together with mismatch-error shaping techniques (MES).

The NS-SAR approach is advantageous because the first several bits can be resolved rapidly using SAR and the remaining bits are resolved using \rDelta\rSigma~modulation over several samples with reduced oversampling-ratio (OSR) to yield a significant overall improvement in conversion efficiency. Reusing the sampling mechanism of the SAR allows the quantisation residue left on V\tss{DAC} to be directly integrated by the loop filter H(z\tps{-1}) that off-sets future conversions and shapes the quantisation noise as 1/(1+H(z\tps{-1})). The main drawback here in comparison to high-resolution \rDS\: modulators is that, while the conversion is faster, the mismatch in the high-resolution DAC must be carefully mitigated. This is where the DWA\cite{sardwa1} and MES\cite{sarmes2} are introduced to eliminate mismatch errors. DWA manipulates the selection of elements used within the MSB capacitive DAC such that the capacitor mismatch is not only decorrelated from the input but is also shaped with a (1-z\tps{-1}) characteristic. The MES module in the LSB section directly off-sets the sampled input using past conversion results to realise a FIR feedback structure such as (1-z\tps{-1}) or (1-2z\tps{-1}+z\tps{-2})~high-pass characteristics to minimise signal-band noise components.

The rest of this paper is organised as follows; Sec.~\ref{Sec:System} will relate the main design parameters to conversion precision in relation to primary noise sources. Once these are established the circuit implementation is presented in Sec.~\ref{Sec:Implementation} together with simulation results in Sec.~\ref{Sec:Results} and Sec.~\ref{Sec:Conclusion} will then conclude this work.

\section{NS-SAR Design} \label{Sec:System}
\begin{figure}
	\centering
	\includegraphics[width=6cm]{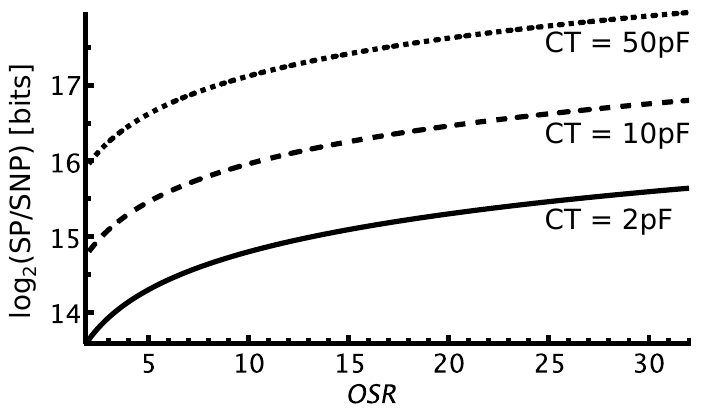}
	\caption{ADC precision as a function of oversampling ratio with respect to SNP while varying sampling capacitance C\tss{T}.}
	\label{Fig:osr_snp}
\end{figure}

\begin{figure}
	\centering
	\includegraphics[width=6cm]{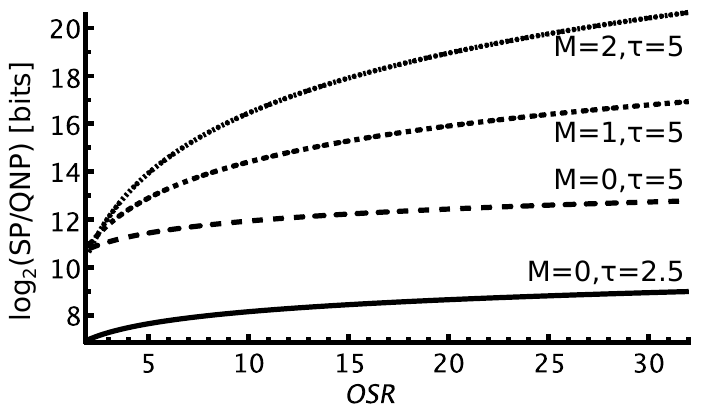}
	\caption{ADC precision as a function of oversampling ratio with respect to QNP while varying settling times $\tau$ and noise-shaping order M.}
	\label{Fig:osr_qnp}
\end{figure}

\begin{figure}
	\centering
	\includegraphics[width=6cm]{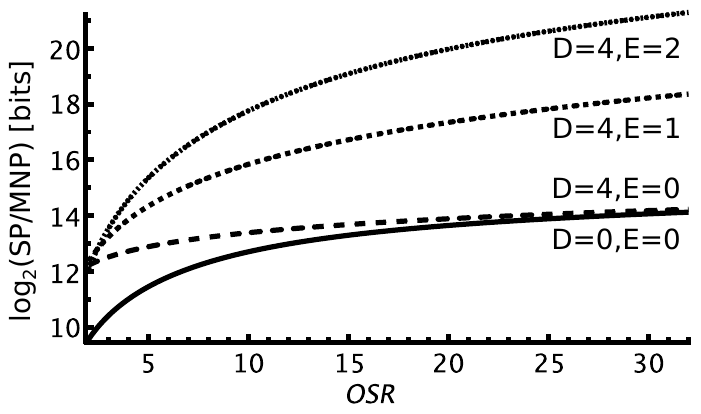}
	\caption{ADC precision as a function of oversampling ratio with respect to MNP while varying calibration D and mismatch-shaping order E.}
	\label{Fig:osr_mnp}
\end{figure}

Comparing with other data-converters, the NS-SAR topology is quite complex with a large number of design parameters that need to be optimised for efficient operation. Below, several of these parameters are discussed in relation to the ADC precision explaining the proposed configuration. Following the single-ended configuration shown in Fig.~\ref{Fig:concept}, we will estimate the expected sampling noise power (SNP), quantisation noise power (QNP), and mismatch noise power (MNP) for the signal bandwidth of fs/(2\,OSR) where fs is the sampling speed. This formulation is purposely presented in brief since it based on established theory from \cite{sdm_book} but it does well to illustrate several trade-off considerations quantitatively when configuring this topology for a particular precision requirement.
\begin{equation} \label{Eq:SNP}
SNP \approx \frac{kT}{C_T} \cdot \frac{2.4}{OSR}
\end{equation}
The expression in Eq.~\ref{Eq:SNP} should be a familiar representation for evaluating the input-referred sampling noise associated with a switched-capacitor integrator. In particular, this corresponds to the input being sampled with a total capacitive value of C\tss{T} using kT as the Boltzman temperature factor. The second term simply arises from averaging the input over OSR cycles together with a correction factor of 2.4 due to the integrator topology in H(z\tps{-1})\cite{sc_noise}. Fig.~\ref{Fig:osr_snp} shows the estimated resolution for several capacitor values assuming we use an input sinusoid with maximum signal power (SP) given a 1.8\,V ADC reference voltage as V\tss{DD}. Inevitably, achieving high resolution implies a large sampling capacitance or a large oversampling ratio. Typically the former is preferred because increasing the capacitive load also decreases the mismatch power from the capacitive DACs.
\begin{equation} \label{Eq:QNP}
QNP \approx \underbrace{\left( V^2_{DD}\,e^{-3\tau} + \frac{V^2_{DD}}{2^{2N}} \right)}_{\text{SAR settling + quantisation }\epsilon} \cdot \frac{\pi^{2M}}{12 \, (1+2M)\,OSR^{1+2M}}
\end{equation}
The expression in Eq.~\ref{Eq:QNP} parametrises the overall SAR resolution as N, the loop fillter order as M, and the number of time constants we allow the capacitive DAC to settle as \rtau \:in order to estimate QNP. This construction shows that settling and quantisation errors are shaped by the loop filter reducing the noise power by the term outside the brackets. Both in Fig.~\ref{Fig:osr_qnp} and in the formulation we observe a strong dependency with regard to M as long as we provide sufficient settling time during SAR conversion. This result suggests that the noise-shaping feed-back must avoid driving the capacitive DAC with active amplifiers during successive-approximation to avoid slowing down the conversion speed or equivalently increasing the power requirement of each amplifier. We can also confirm here that the order of the loop filter does not need to be very high if the QNP needs to match the SNP.
\begin{equation} \label{Eq:MNP}
MNP \approx \underbrace{\left( \frac{\pi^2\,2^{-2D}}{3 \cdot 2^K \, OSR^3} + \frac{\pi^{2E}\,2^{-2K}}{(1+2E) OSR^{1+2E}} \right)}_{\text{DWA-MSB + MES-LSB DAC}} \frac{\sigma^2 \, V^2_{DD}}{3}
\end{equation}
The MNP is evaluated in Eq.~\ref{Eq:MNP} with respect to the MES noise shaping order E, the number of bits D used to calibrate each capacitor in the MSB DAC in an idealised way. K represents the MSB DAC resolution in bits. Using a capacitor standard deviation $\sigma=0.5\%$ and K=4, the MNP of several configurations is shown in Fig.~\ref{Fig:osr_mnp}. The observation here is that for small OSR values the mismatch noise is typically dominated by the MSB DAC as the mismatch is not sufficiently shaped. It is relatively expensive to increase the number of elements in the MSB DAC since the scaling is linear and increasing the OSR diminishes the advantage of performing SAR. Instead we propose to calibrate the 15 capacitors in the MSB section as D will reduce the MNP more efficiently. The mismatch from the LSB section contains many more elements and is more effectively shaped using a second-order MES technique.

The above trends are used to optimise the FOM\tss{S} in a similar fashion to \cite{nssar4} by correlating hardware requirements with power and accuracy estimators for several configurations. Given an initial 18\,bit target precision, we propose the following configuration: CT=50\,pF, M=2,\rtau=5, K=5, D=4, E=2 with the OSR set to 16 to ease the decimation effort.

\section{Circuit Implementation} \label{Sec:Implementation}

\begin{figure}
	\centering
	\includegraphics[width=7.5cm]{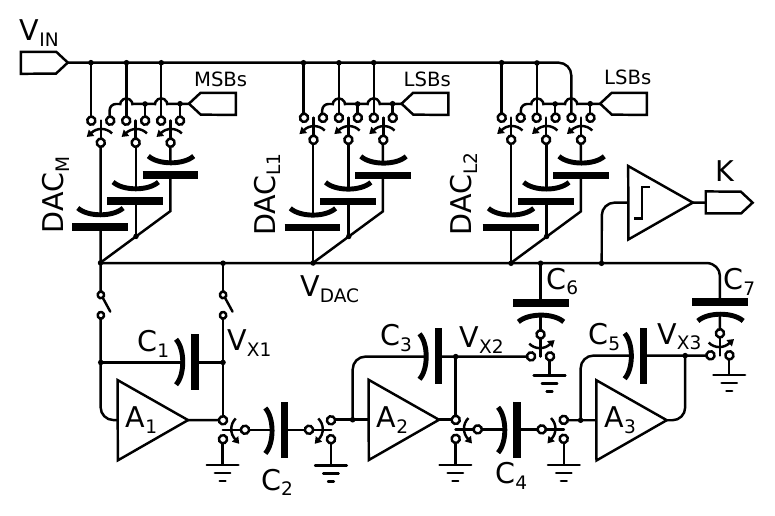}
	\caption{Implementation of the capacitor network used to perform signal conversion using the bottom sampled capacitor arrays DAC\tss{M} for the DWA bits and DAC\tss{L1} \& DAC\tss{L2} for the MES bits. The loop filter is also shown where A\tss{1} amplifies the quantisation residue that is then integrated by A\tss{2} \& A\tss{3} for noise-shaping.}
	\label{Fig:cdac}
\end{figure}

The analogue part of the ADC implementation is shown in Fig.~\ref{Fig:cdac}. Note that the implemented ADC uses an equivalent fully-differential configuration to gain extra input-dynamic range as well as digital noise suppression. This realisation is entirely based on manipulating the capacitive DAC and enables low-power operation for varying sampling rates. A second distinguishing feature of the proposed topology is that the comparator only requires one input terminal opposed to two seen in prior-art \cite{nssar0,nssar1} which leads to better linearity and noise performance. In addition the input is bottom plate sampled such that sensitivity to parasitic capacitance and comparator non-linearity is considerably reduced. This figure also shows three capacitor arrays where the DAC\tss{M} section corresponds to the DWA modulated MSBs and the DAC\tss{L1/L2} section represents the MES modulated LSBs being fed back from the SAR controller. Implementing the second-order MES noise-shaping uses the ping-pong configuration from \cite{sarmes1}.

Three switched-capacitor amplifiers are used to realise a second-order cascaded-feed-forward-integrator (CFFI) loop filter topology where the first stage provides auto-zeroing as well as signal amplification by $C_T/C_1\approx30$. This design uses an asynchronous SAR conversion process~\cite{asyncsar} which is why there are only 3 phases in the switched capacitor circuit; the sampling phase (SMP), the successive approximation phase (SAR), and the quantisation filtering phase (QNF). The SAR only takes 100\,ns and the FSM immediately initiates the QNF phase reducing the input clock to twice the sampling rate. The three phases operate as follows:

\begin{enumerate}
	\item[\textbf{SMP}] First A\tss{1} actively samples its offset on the top plate while bottom plate samples V\tss{IN} on DAC\tss{M} together with the MES code on DAC\tss{L1/L2}. A\tss{2/3} are simultaneously integrating quantisation errors and sampling the result V\tss{X2/X3} with respect to V\tss{DAC} on C\tss{6} and C\tss{7}.
	\item[\textbf{SAR}] V\tss{DAC} then converges to virtual ground by switching the input to DAC\tss{M/L1/L2} while quantisation errors from prior conversions are removed by grounding the bottom plate of C\tss{6/7}. This also disconnects A\tss{1/2/3} from V\tss{DAC}.
	\item[\textbf{QNF}] Finally DAC\tss{M/L1/L2} is held and the resulting quantisation residue left on V\tss{DAC} is amplified by A{1} on V\tss{X1}. C\tss{2/4} samples the voltages V\tss{X1/X2} which are used to integrate during the following SMP phase.
\end{enumerate}

This configuration scales well for varying loop filter structures as 80\% of the power is dissipated by A\tss{1} and the total sampling noise is dominated by C\tss{T}. The comparator uses a conventional strong-arm topology that is carefully designed to minimise off-set since this off-set will be seen at the output of A\tss{3} after amplification which can diminish the output-swing. Conversely the noise and distortion characteristics of the analogue filtering chain is proportionally reduced when the signal is fed back onto the capacitor array during sampling as the attenuation ratio $C_{6-7}/C_T$ inverts the amplification ratio with good matching.


The MSB DAC calibration mechanism is uses a digital shuffling technique to identify mismatch by switching out different sets of capacitors that will only incur voltage fluctuation on V\tss{DAC} in the presence of mismatch\cite{calisar1}. These errors are then amplified by A\tss{1} after the SAR \& QNF process and digitally tunes each MSB capacitor using a capacitive sub-DAC. The sign of each shuffling result is accumulated to adjust the the 15 calibration codes thereby eliminating the mismatch in the MSB DAC. This process can be performed in the background without requirements on the input signal because DWA randomises the capacitor selection mechanism during shuffling.

\begin{table*}
	\caption{Performance summary and comparison with state of the art}
	\label{TB:Summary}
	\centering
	\begin{tabular}{p{1.8cm} p{1.2cm} p{1.2cm} p{1.2cm} p{1.2cm} p{1.2cm} p{1.2cm} p{1.2cm} p{1.2cm}}
		\toprule
		Spec.       	& This Work & \cite{nssar2} & \cite{calisar1} & \cite{nssar5} & \cite{nssar6} & \cite{nssar4} & \cite{nssar0} & \cite{nssar1} \\
		\midrule
		Year									& 2018						&  2018				& 2018 				& 2018 				& 2018 				& 2017 				& 2016 				&	2012 \\
		Tech.\hfill[nm] 			& 180 						&  180				& 180					& 28	 				& 40	 				& 180	 				& 55  				&	65	\\
		Supply\hfill[V]      	& 1.8    					&  1.8				& 1.8/5				& 1.1/1.2			& 2.5/1.1	 		& 1.2 				& 1.2  				&	1.2	\\
		Power\hfill[W]      	& 68\rmu  				&  7.93\rmu		& 12.9m				& 4.2m 				& 140\rmu			& 5.16\rmu		& 15.7\rmu		&	806\rmu	\\
		Topology 		 					& NS-SAR 					& \rDS-SAR		& SAR 				& CT-\rDS			& \rDS-SAR		& \rDS-SAR		& NS-SAR			&	NS-SAR	\\
		DAC Res.\hfill[b] 		& 10 							&  9 					& 20 					& 4		 				& 7		 				& 8		 				& 12					&	8	\\
		NS-Order 			 				& 2$^\dagger$			&  1					& 0 					& 2$^\dagger$	& 3		 				& 2		 				& 1$^\dagger$	&	1$^\dagger$	\\
		OSR 			 						& 16 							&  256    		& 1 					& 16	 				& 12	 				& 24	 				& 256    			&	4	\\
		BW\hfill[Hz] 					& 15.6k  					&  1k  				& 500k				& 10M	 				& 40k	 				& 100k 				& 4k					&	11M	\\
		SNDR\hfill[dB]      	& 102		 					&  85					& 102					& 94	 				& 84					& 67	 				& 96.1 				&	62	\\
		Area\hfill[mm\tsqrd] 	& 0.201						&  0.68				& 4 					& 0.1	 				& 0.07 				& 0.02 				& 0.07				&	0.03	\\
		FoM\tss{S}\hfill[dB]  & 183$^\star$ 		&  166				& 176					& 168	 				& 169	 				& 170	 				& 180					&	164	\\
		\bottomrule
	\end{tabular}
	\flushleft{\thanks{$^\star$ Estimated based on post-layout simulation results where FoM\tss{S} = SNDR + 10log\tss{10}(BW/P). $^\dagger$ FIR \& digital noise-coupling poles excluded.}}
\end{table*}%

\section{Simulation Results} \label{Sec:Results}
\begin{figure}
	\centering
	\includegraphics[width=7.5cm]{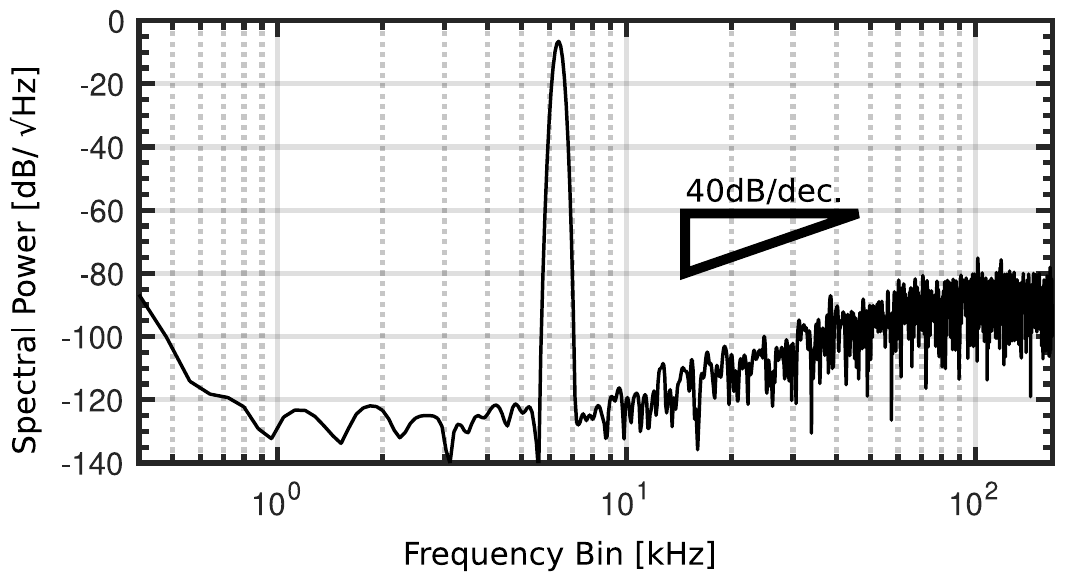}
	\caption{Post-layout simulation result showing the noise-shaped output spectrum from a -3\,dBFS  input sinusoid at 6.5\,kHz.}
	\label{Fig:sim_thd}
	\vspace{-5mm}
\end{figure}

The proposed NS-SAR has been designed and validated using a commercially available 180\,nm TSMC technology (1P6M HV BCD GEN II). All sub-circuits have been integrated with reconfigurable \rDelta\rSigma, DWA, MES, and calibration modes to fully characterise post-silicon performance that will confirm the evaluation in Sec.~\ref{Sec:System}. This circuit uses an analogue and digital supply at 1.8\,V, a 1\,\rmu A current reference to bias A\tss{1-3}, and a 0.9\,V common-mode reference for V\tss{CM}-based capacitor switching. Preliminary post-layout simulation results are shown in Fig.~\ref{Fig:sim_thd}. This demonstrates the ADC can resolve 17\,bits of precision without distortion while using an external clock of 1\,MHz where one cycle is used to sample the input and one cycle is used for conversion plus quantisation noise shaping and another cycle is optionally used for background calibration. The last phase can be skipped if the MSB capacitors are already tuned to speed-up signal conversion to 31.25\,kS/s since temperature and voltage variations over time during normal operation will typically not corrupt the calibrated capacitor characteristics.

\begin{figure}
	\centering
	\includegraphics[width=6.5cm]{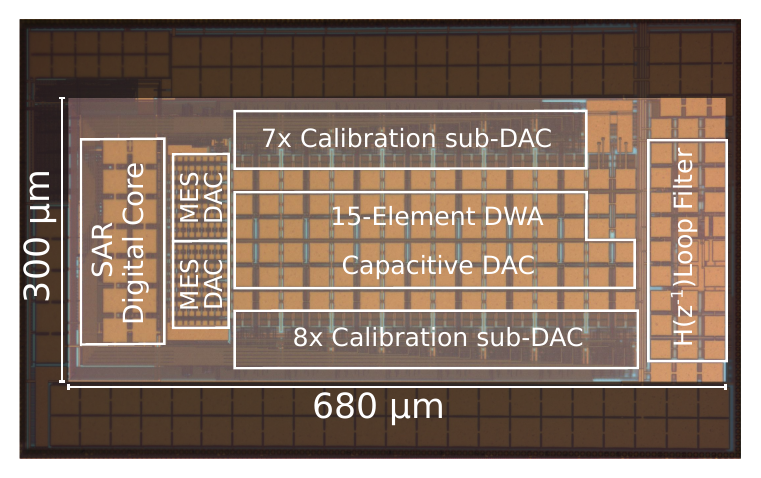}
	\caption{ADC micro-photograph showing labelled blocks in relation to Fig~\ref{Fig:concept} where the MES and DWA circuitry is included in the main digital core. Decoupling capacitors are placed over active circuitry or underneath active mim-caps.}
	\label{Fig:layout}
	\vspace{-5mm}
\end{figure}

The layout for this ADC is shown in Fig.~\ref{Fig:layout}. A large majority of silicon area is dedicated towards the MSB capacitive array as the sampling noise must be suppressed. The switched capacitor integrator can be relatively small because the internal loop-filter gain reduces its sampling noise. The digital core takes up a considerable amount of area and power budget primarily as a result of using a 180\,nm CMOS technology where more advanced technologies may lead to further improvements if the 1.8\,V rating can be maintained. Each MSB capacitor is trimmed using a 8\,bit sub-DAC that tunes about 5\% of the 1.7\,pF unit capacitance which accommodates well over 3\rsigma~of the expected capacitor mismatch as well as wafer level variations that may not be captured by the typical mismatch model. The performance measures for the proposed ADC are shown in Table~\ref{TB:Summary}. Again we highlight the fact that while all these works have highly optimised power budgets, this topology is able to achieve over 100\,dB SNDR with a 10$\times$ lower oversampling ratio than prior art for this level precision. While this does imply a marginally increased area requirement, the peak efficiency can be achieved over a greater span of sampling frequencies. Note that this particular TSMC process kit does not allow post-layout Monte-Carlo so the calibration will be validated using post-silicon results.

\section{Acknowledgement} \label{Sec:Acknowledgement}

This work was supported by the UK Engineering and Physical Sciences Research Council (EPSRC) grants EP/M020975/1 \& EP/R024642/1.

\section{Conclusion} \label{Sec:Conclusion}

This works presents a 17\,bit Noise Shaping SAR ADC with reduced oversampling ratio and a purely capacitive implementation which enables in state-of-the-art conversion efficiency over a large range of sampling frequencies. In comparison with conventional over-sampling ADCs simulation results suggest this NS-SAR is able to achieve 102\,dB SNDR with substantially lower noise-shaping requirements with comparable or reduced circuit complexity while achieving better power efficiency. We also demonstrated a high-level parameter selection methodology that is used to optimise the FoM\tss{S} and identify the factors limiting ADC precision.\\

\input{main.bbl}

\end{document}